\documentclass[a4paper,aps,prb,twocolumn,floatfix,showpacs,superscriptaddress,footnoteinbib]{revtex4-1}
\usepackage{graphics}
\usepackage{epsfig}
\usepackage{lipsum, babel}
\usepackage{times}
\usepackage{xcolor}   
\usepackage{amsfonts}
\usepackage{amsmath}
\usepackage{amssymb}
\usepackage{amsthm}
\usepackage{hyperref} 
\usepackage{wasysym}
\usepackage{bm} 
\usepackage{graphicx}
\newcommand{\ie}{{\it i}.{\it e}.,~}
\newcommand{\eg}{{\it e}.{\it g}.,~}

\DeclareMathOperator{\sech}{sech}

\begin{document}

\title{Effects of surface potentials on Goos-H\"{a}nchen and Imbert-Fedorov shifts in Weyl semimetals}

\author{Ninad Kailas Dongre}
\affiliation{Department of Physics, Indian Institute of Technology (BHU), Varanasi 221005, India}
\affiliation{Department of Physics, The University of Texas at Dallas, Richardson, TX 75080, USA}
\author{Krishanu Roychowdhury}
\affiliation{Department of Physics, Stockholm University, SE-106 91 Stockholm, Sweden}

\begin{abstract}
Weyl semimetals exhibit exotic transport responses, among which, recently Goos-H\"{a}nchen (GH) and Imbert-Fedorov (IF) effects have received a revived attention, which are, otherwise, well-studied phenomena in optical systems and certain electronic systems. Besides the usual parametric dependence of the shifts inherited from the underlying Hamiltonian to describe the Weyl system and/or that induced by external controls, the IF shift further carries a topological identity -- it depends on the chirality of the Weyl cones. A realistic system of Weyl semimetal naturally accommodates surface potentials induced by impurities present on its surface that could pose impediments to observe clean transport signatures predicted in theoretical models. Classifying these potentials, we study their effects on GH and IF shifts to provide useful guidance to future experiments that are tuned to the objective of characterizing Weyl semimetals and for a possible realisation of novel devices  based on these phenomena. A transfer matrix-based approach is invoked to study the profile of Weyl wavefunctions across the interface which is hosting the impurity potentials, revealing that such potentials can lead to several discerning effects which, in certain cases, extend even to nullifying the IF shift completely and giving rise to phenomenon like valley inversion.     
\end{abstract}

\maketitle

\section{Introduction}

The Goos-H\"{a}nchen~\cite{goos1947neuer, artmann1948berechnung, mcguirk1977angular} (GH) and Imbert-Fedorov~\cite{fedorov1955k, schilling1965strahlversetzung, imbert1972calculation} (IF) shifts, first discovered in the context of wave-optics, have now found application in a variety of systems~\cite{wild1982goos, bretenaker1992direct, pfleghaar1993quantitative, emile1995measurement, bonnet2001measurement, berman2002goos, li2003negative, fan2003amplified, shadrivov2003giant, felbacq2003goos, yin2004large, merano2007observation, beenakker2009quantum, de2010observation, nieminen2020goos} that include optical waveguides, metamaterials, plasmonics, and quantum systems. These effects transpire for incident beams of finite width (\ie a distributed spectrum of plane waves) whose reflection and refraction do not quite follow the simple geometric rules of Newtonian optics feature lateral and angular shifts with respect to the point of incidence instead. For instance, when a traveling photon beam suffers multiple total internal reflections inside an optical waveguide, measurable shifts are observed that may be attributed to a finite penetration of the evanescent beam into the cladding material at each turn. This is reminiscent of the tunneling of quantum particles through finite barriers which is accredited to their dual (wave-particle) nature and has inspired researchers to explore these scattering effects in quantum systems such as two-dimensional electron gas (2DEG) nanostructures~\cite{chen2008tunable} (for relativistic corrections, see Ref.~\onlinecite{miller1972shifts, fradkin1974spatial}) and Dirac materials like graphene~\cite{beenakker2009quantum, chen2011goos, sharma2011electron, zhai2011valley, cheng2012goos, wang2013guided, chen2013electronic, wang2013resonant, grosche2015goos, zeng2017tunable, wu2018transitional} and transition metal dichalcogenides~\cite{das2021quantum}, enabling new device applications that extend, as well, to terahertz regime~\cite{fan2016electrically, zheng2019enhanced, liu2020high}. The GH shift has been exploited in a system of 2DEG subject to a tunable electric potential and hosted between magnetic stripes to construct an efficient spin beam splitter~\cite{chen2008tunable, zhang2014spin}. While spin manipulation of such type is highly desirable for spintronics and quantum information applications, the valley degrees of freedom in electronic systems (such as semiconductors and semimetals) are also of concurrent and intensive interest~\cite{zhai2011valley, wu2011valley}, especially for emerging quantum technologies like valleytronics. 

On both fronts, Dirac materials have enticed specific attention over the last few decades. Graphene (along with other materials of similar band structures), among them, has emerged as a paradigmatic model wherein a myriad of exotic electronic phenomena that arise in two dimensions have been proposed and verified in experiments (see Ref.~\onlinecite{peres2010colloquium} for a review). The GH shift for the massless electrons in graphene manifests as a pseudospin-dependent scattering effect that results in a quantized jump in the conductance of heterojunctions~\cite{beenakker2009quantum}. In the same system, a valley-dependent GH shift is studied by means of tailoring the local strain profile~\cite{wu2011valley} exhibiting a close resemblance to the spin manipulation in Ref.~\onlinecite{chen2008tunable} and also, in part, the aforementioned optical and electronic analogs.  

The IF effect, likewise, has also been explored in graphene systems~\cite{kort2016quantized, farmani2017tunable, xu2017active, luo2018magneto, zhu2019electro}, and recently resuscitated in Weyl semimetals~\cite{jiang2015topological, jiang2016chiral, wang2017imbert, hao2019imbert, ye2019goos, chattopadhyay2019fermi, liu2020lattice}. The latter is characterized with topologically robust nodes (immune to arbitrary perturbations) in the bulk and Fermi arc surface states~\cite{hosur2013recent}. The nodes are of distinct chirality, referred to as Weyl cones (singly degenerate, as opposed to the doubly degenerate Dirac cones in, \eg graphene), and responsible for strange phenomena like {\it chiral anomaly}, exclusive in three dimensions. Aside from the bulk nodes, the surface states are also a distinctive hallmark of these semimetals which have been experimentally probed and bear important implications for various transport properties of these systems such as TaAs or NbAs (see Ref.~\onlinecite{hasan2017discovery} and references therein). 

In realistic materials, surfaces of a Weyl semimetal would naturally host impurities that give rise to different types of surface potentials, akin to a common scenario in topological insulators~\cite{zhang2012surface, roy2016transport, roy2016pseudospin}. These potentials could break time-reversal or inversion (or none), which is anyway required to stabilize a Weyl semimetallic phase in the bulk. For instance, magnetic impurities can populate the surface of a magnetic Weyl semimetal such as Co$_3$Sn$_2$S$_2$~\cite{liu2019magnetic} while spin-orbit type impurities can be found in inversion breaking Weyl semimetals such as candidates from the TaAs material class,~\cite{huang2015inversion}. Apart from bulk impurities, these potentials are also of great relevance to experiments dedicated to probing surface transport (such as conductivities) in these topological materials and also in part attesting the theoretical predictions based on clean surfaces such as those relying on the bulk-boundary correspondence. 

Among many intriguing consequences, localized bulk impurities in Weyl semimetals can lead to suppression in the nodal density of states which may or may not be lifted by impurity-induced resonances~\cite{huang2013impurity}. Besides, these impurities also exhibit prominent features for the surface states such as giving rise to bound states for certain parameter values in the Hamiltonian that lead to distinct topological phases~\cite{he2018impurity}. As the lateral shifts concerned are exclusively surface phenomena and that such shifts, particularly for a Weyl semimetal surface, are intricately related to the Fermi arc structure~\cite{chattopadhyay2019fermi}, we naturally ask to what extent are they influenced by the presence of impurity-induced surface potentials which are compatible with the symmetry breaking in the bulk of a Weyl system. The central result of this article engages a thorough analysis of GH and IF shifts in presence of such imperfections, which have hitherto been studied only for clean surfaces.

The remainder of the article is structured as follows- In Sec.~\ref{sectwo}, we introduce the model and review some aspects of the GH and the IF shift for a clean surface. In Sec.~\ref{secthree}, we discuss the transfer matrix based approach that applies to analyze the Weyl wavefunction across an interface hosting impurity-induced surface potentials. These potentials are classified in Sec.~\ref{secfour} where the shift calculations are revisited for each of the classes. We summarize the findings in Sec.~\ref{secfive} and discuss future work. 

\section{GH and IF shifts for a clean interface}\label{sectwo}

\begin{figure}[t]
\begin{center}
  \includegraphics[width=1.0\linewidth]{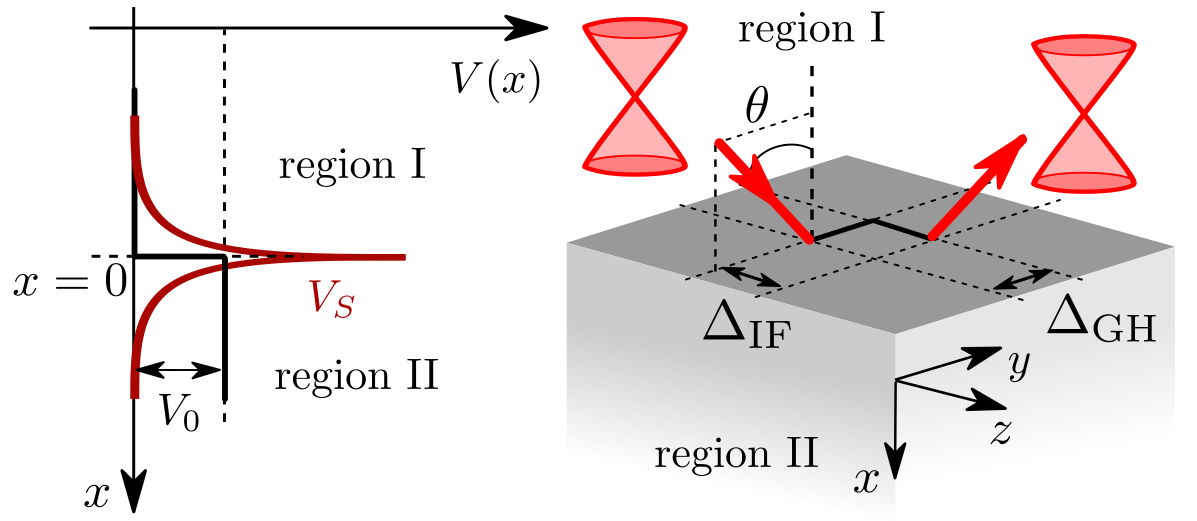}
\end{center}\vspace{-4mm}
\caption{The schematic of the Weyl semimetal surface hosting surface potential at $x=0$ with the two lateral shifts indicated. The plane of incidence is the $x-y$ plane such that in-plane lateral shift (GH shift) is along the $y$-axis and the out-of-plane shift (IF shift) is along the $z$-axis.  
}
\label{fig:fig1}
\end{figure}

We start by reviewing the results from a  previous work that computed the GH and the IF shifts for a clean interface~\cite{jiang2015topological}. 
For calculations, we consider the surface (or interface) to be the $y-z$ plane at $x=0$ which breaks the translation symmetry along $x$ but retains along the other two directions (see Fig.~\ref{fig:fig1}), however, the formulation presented in this article, as well, applies to surfaces of other orientations.

The Hamiltonian of the system is described by
\begin{align}
 H=
 \begin{cases}
  H_{\rm I}=\chi\hbar v_F ~ k_{i}\sigma_{i}  & x \leq  0\\
  H_{\rm II}=\chi\hbar v_F ~ k_{i}\sigma_{i}+V_0 & x> 0,
 \end{cases}
 \label{ham}
\end{align}
where $i\in\{x,y,z\}$ and $\chi$ ($\chi=\pm1$) denotes the chirality of the Weyl cones. We refer to the region $x\leq0$ as region I, and the region $x>0$ as region II as in Fig.~\ref{fig:fig1}. These two regions are distinguished by applying a chemical potential difference $V_0$ between them which creates a finite barrier for the Weyl fermions to scatter off. Further, $v_F$ denotes the Fermi velocity, which is taken the same on both sides, and $\sigma_{i}$ are the Pauli matrices representing the pseudospin degrees of freedom. Throughout the calculations that follow, we adopt the units in which $\hbar=v_F=1$. A beam of Weyl fermions, in the form of a Gaussian wavepacket, incident on the interface at $x=0$ is modelled by 
\begin{align}\label{incidwvpckt}
 \Psi^{\rm in}(\mathbf{r})=\int^{\infty}_{-\infty}{\rm d}k_{y} {\rm d}k_{z}~f(k_y, k_z)\psi^{\rm in}(\mathbf{k},\mathbf{r}),
\end{align}
where the Gaussian function $f$ is assumed to be localized around $(\Bar{k}_{y}, \Bar{k}_{z})$ as
\begin{align}
 f(k_y, k_z)= \frac{1}{2\pi\sqrt{\Delta_{y}\Delta_{z}}}{\rm Exp}\bigg[{-\frac{(k_{y}-\Bar{k}_{y})^2}{2\Delta_{y}^{2}}-\frac{(k_{z}-\Bar{k}_{z})^2}{2\Delta_{z}^{2}}}\bigg],
\end{align}
$\Delta_{y}$ ($\Delta_{z}$) denoting the width of the wavepacket along $y$ ($z$). The spinor part of the incident wavefunction $\psi^{\rm in}$ satisfies the Schr\"{o}dinger equation $H_{\rm I}\psi^{\rm in}=E\psi^{\rm in}$, and including the plane wave phase factor $e^{i{\bf k}\cdot{\bf r}}$,
\begin{align}
 \psi^{\rm in}(\mathbf{k},\mathbf{r}) = C_i
 \begin{bmatrix} 
 2\eta\chi e^{-i\theta/2} \\ \tilde{\eta} e^{i\theta/2} 
 \end{bmatrix} e^{i{\bf k}\cdot{\bf r}},
 \label{incidspinor}
\end{align}
where $k_x = \sqrt{E^{2}-k_{y}^{2}-k_{z}^{2}}$, $\tan\theta=k_y/k_x$, $\eta=\sqrt{\frac{E-k_z}{E+k_z}}$, $\tilde{\eta}=\eta^2(1+\chi)+(1-\chi)$, and $C_i$ is the normalization constant. Taking the plane of incidence to be the $x-y$ plane, $\theta$ represents the angle of incidence for the incident beam measured from the surface normal perpendicular to the $y-z$ plane at the point of incidence (see Fig.~\ref{fig:fig1}).

The reflected wavepacket can be expressed in a similar form as the incident one. Multiplied by the reflection coefficient $r=|r|e^{i\phi_r}$, it is
\begin{align}\label{rewvpckt}
 \Psi^{\rm re}(\mathbf{r})=\int^{\infty}_{-\infty}{\rm d}k_{y} {\rm d}k_{z}~f(k_y, k_z)~r\psi^{\rm re}(\mathbf{k},\mathbf{r}),
 \end{align}
where $\phi_r$ is the reflection phase and $\psi^{\rm re}(\mathbf{k},\mathbf{r})$ is obtained from $\psi^{\rm in}(\mathbf{k},\mathbf{r})$ in Eq.~\ref{incidspinor} via $k_x\rightarrow -k_x, \theta\rightarrow\pi-\theta$. For the phenomena concerned, we will be considering total reflection from the interface in region I with $|r|=1$, and for this, the mode on the other side (region II) must be evanescent. Note this happens only for values of the incident angle $\theta$ greater than a critical value $\theta_c$ which, as will be shown later, depends on the ratio of the barrier height to the incident energy $V_0/E$. 

Linearizing the phases $\theta$ and $\phi_r$ in terms of $\Bar{k}_y$ and $\Bar{k}_z$, the integrals in Eq.~\ref{incidwvpckt} and \ref{rewvpckt} provide the centers of the incident and the reflected wavepackets respectively from which the (spinor) component-wise spatial shifts along $y$ and $z$ follow as
\begin{align}
 \Delta^{y}_{\pm} &= -\frac{\partial}{\partial k_{y}}\phi_{r}(\Bar{k}_{y},\Bar{k}_{z})\mp\frac{\partial}{\partial k_{y}}\theta(\Bar{k}_{y},\Bar{k}_{z}), \nonumber \\
 \Delta^{z}_{\pm} &= -\frac{\partial}{\partial k_{z}}\phi_{r}(\Bar{k}_{y},\Bar{k}_{z})\mp\frac{\partial}{\partial k_{z}}\theta(\Bar{k}_{y},\Bar{k}_{z}),
 \end{align}
where, $\pm$ refer to the shifts of the two spinor components. The full spatial  shifts are then given as the weighted average of the individual shifts for each of the spinor components,
\begin{align}\label{shift1}
 \Delta^{y(z)}=\frac{4\eta^2\Delta^{y(z)}_{+}+\tilde{\eta}^{2}\Delta^{y(z)}_{-}}{4\eta^2+\tilde{\eta}^2}.
 \end{align}\
As the results do not depend on the shape of the wavepackets, we conveniently adopt a reference by setting $\Bar{k}_z=0$ which aligns the GH and the IF shift along $y$ and $z$-axis respectively and, thus, simplifies Eq.~\ref{shift1} to
\begin{align}\label{shift2}
 \Delta_{\rm GH} &\equiv \Delta^{y}=\frac{4\eta^2\Delta^{y}_{+}+\tilde{\eta}^{2}\Delta^{y}_{-}}{4\eta^2\chi^2+\tilde{\eta}^2}=-\frac{\partial \phi_{r}}{\partial k_{y}}, \nonumber \\
 \Delta_{\rm IF} &\equiv \Delta^{z}=\frac{4\eta^2\Delta^{z}_{+}+\tilde{\eta}^{2}\Delta^{z}_{-}}{4\eta^2\chi^2+\tilde{\eta}^2}=-\frac{\partial \phi_{r}}{\partial k_{z}}.
\end{align}
To quantify the shifts, it then remains to compute the reflection phase $\phi_r$ using the continuity of the wavefunctions at the interface
\begin{align}\label{conti}
 \psi^{\rm tr}(\mathbf{k},0) = \psi^{\rm in}(\mathbf{k},0) + \psi^{\rm re}(\mathbf{k},0) 
\end{align}
where the transmitted spinor  $\psi^{\rm tr}(\mathbf{k},\mathbf{r})$ in region II has an evanescent form at energy $E$ as
\begin{align}
 \psi^{\rm tr}(\mathbf{k},\mathbf{r}) = C_t
 \begin{bmatrix} 
 -i \\ \beta_\chi 
 \end{bmatrix} e^{-\kappa x}e^{i(k_y y + k_z z)};~\beta_\chi=\frac{k_{y}+\kappa}{k_{z}+\chi(E-V_0)},
 \label{transmitspinor}
\end{align}
$C_t$ being the normalization constant and $\kappa =\sqrt{k_{z}^{2}+k_{y}^{2}-(E-V_{0})^{2}}>0$ setting the inverse decay length. Eq.~\ref{conti} yields a chirality-dependent reflection phase
\begin{align}
 \phi_{r}=-\theta-\frac{\pi}{2}+2\tan^{-1}\Big[\frac{\tilde{\eta}\cos\theta}{2\eta\chi\beta_{\chi}-\tilde{\eta}\sin\theta}\Big],
\end{align}
from which the spatial shifts follow as
\begin{align}
 \Delta_{\rm GH} = \frac{(1+\sin^{2}{\bar\theta}-\frac{V_0}{E})}{\kappa\sin{\bar\theta}\cos{\bar\theta}} ~~;~~\Delta_{\rm IF} = -\frac{\chi}{E\tan{\bar\theta}},
 \label{shiftclean}
\end{align}
where $\tan{\bar\theta}={\bar k}_y/\bar{k}_x$. The critical angle to ensure a total reflection from the interface is given by $\theta_c=\sin^{-1}|V_0/E-1|$. 

Note the GH shift on a clean surface is not a chirality dependent phenomenon while the IF shift is and can be interpreted as a topological effect~\cite{jiang2015topological}. This is attributed to the fact that during the reflection, the Weyl fermions retain their valley characteristics due to momentum conservation on the surface which is in sharp distinction with the optical analog where the polarization of the photons does get altered during reflection. We further note that the IF shift is independent of the ratio of the barrier height to the incident energy $V_0/E$. The GH shift, on the other hand, changes sign at a given angle of incidence $\theta^\ast=\sin^{-1}\sqrt{\sin\theta_c}$ -- it is negative for $\theta_c<\bar{\theta}<\theta^\ast$ while positive for $\theta^\ast<\bar{\theta}$ irrespective of the values of $V_0/E$, and likewise for $\bar{\theta}\rightarrow-\bar{\theta}$. 

So far, we have discussed the shifts for a clean interface. In the following sections, we will demonstrate how they get modified when the same interface harbors different types of impurity-induced surface potentials that affect the continuity equation stated above and the resultant reflection phase. In fact, in the presence of certain types of surface potentials, the shifts feature distinct asymmetry between the valleys resembling the effects of intervalley scattering~\cite{wang2017imbert}.

\section{The transfer matrix approach}\label{secthree}

For Weyl (or Dirac) fermions subject to a delta potential scattering, the transfer matrix accounts for the rotation between the spinors on the two sides of the potential~\cite{calkin1987proper}. This approach is adopted extensively in transport calculations to compute observables like surface conductance, Aharonov-Bohm oscillations, spin Berry phase in various mesoscopic systems including topological insulators with impurities \cite{ilan2012nonequilibrium, xypakis2020perfect, adak2020spin}. 

The spatial profile of the full Hamiltonian in presence of a delta function-type surface potential $V_S(x)={\cal V}\delta(x)$ is given by
\begin{align}\label{fullHam}
 {\cal H} = \chi(\sigma_a k_a-i\sigma_x\partial_x) + V_0\Theta(x) + {\cal V}\delta(x),
\end{align}
where $a\in\{y,z\}$ as $k_{y,z}$ are regarded good quantum numbers for we have imposed periodic boundary conditions on the interface; ${\cal V}$ denotes the surface potential [see Fig.~\ref{fig:fig1}] which could be a scalar or a spin-valued operator. The Schr\"{o}dinger equation ${\cal H}\psi=E\psi$ can be recast as $\partial_x\psi={\cal H}_0(x)\psi$ leading to a path-ordered solution of $\psi(x)$ as $\psi(x_2)={\cal T}_{x_2,x_1}\psi(x_1)$ where the transfer matrix ${\cal T}={\cal P}_x{\rm Exp}\big[\int_{x_1}^{x_2}{\rm d}x~{\cal H}_0(x)\big]$, ${\cal P}_x$ denoting the path-ordering~\cite{adak2020spin}. For the interface at $x=0$ (see Fig.~\ref{fig:fig1}), the transfer matrix pertinent to the model in Eq.~\ref{fullHam} turns out to be
\begin{align}
 {\cal T} = \lim_{\epsilon\to0}{\cal P}_x~ {\rm Exp}\bigg[-i\int_{-\epsilon}^{+\epsilon}{\rm d}x~ \sigma_x V_S(x) \bigg]=e^{-i\sigma_x {\cal V}}.
\end{align}
The continuity equation in Eq.~\ref{conti} gets modified to 
\begin{align}\label{conti2}
 \psi^{\rm tr}(\mathbf{k},0^+) = {\cal T} [\psi^{\rm in}(\mathbf{k},0^-) + \psi^{\rm re}(\mathbf{k},0^-)],
\end{align}
and so does the reflection phase $\phi_r$ accordingly. This is how the spatial shifts are expected to get modified in presence of various surface potentials.

\section{Surface potentials}\label{secfour}

In this section, we will consider a family of surface potentials by expressing the surface term ${\cal V}$ in Eq.~\ref{fullHam} as
\begin{align}
 {\cal V} = \lambda + {\bf V}\cdot{\bf \sigma},
\end{align}
where $\lambda$ is a real constant, and the components of the vector ${\bf V}=(V_x, V_y, V_z)$ are, in general, real functions of $k_y$ and $k_z$. In particular, we take note of the following cases:
\begin{itemize}
 \item {\bf Scalar potential:} $\lambda\neq 0$, ${\bf V}=0$.
 \item {\bf Magnetic potential:} Uniform magnetic field specified by $V_x=B_x$, $V_y=B_y$, $V_z=B_z$ where $B_{x,y,z}$ are constants and $\lambda=0$.
 \item {\bf Spin-orbit potential of Rashba type:} $\lambda=V_x=0$, but $V_y$ and $V_z$ are linear functions of $k_y$ and $k_z$ as $V_y=-\alpha_R k_z$ and $V_z=\alpha_R k_y$.
 \item {\bf Spin-orbit potential of Dresselhaus type:} $\lambda=V_x=0$, but $V_y$ and $V_z$ are linear functions of $k_y$ and $k_z$ as $V_y=\alpha_D k_y$ and $V_z=-\alpha_D k_z$.
\end{itemize}
In the following subsections, we compute the GH and the IF shift from the resultant $\phi_r$ while addressing the above four cases separately. It should, however, be noted that combinations of them are also likely to occur on the surface of a Weyl semimetal. We will illustrate one such combination in which the surface potential includes both scalar and magnetic impurities.


\subsection{Scalar potential}

\begin{figure}[t]
\begin{center}
  \includegraphics[width=1.0\linewidth]{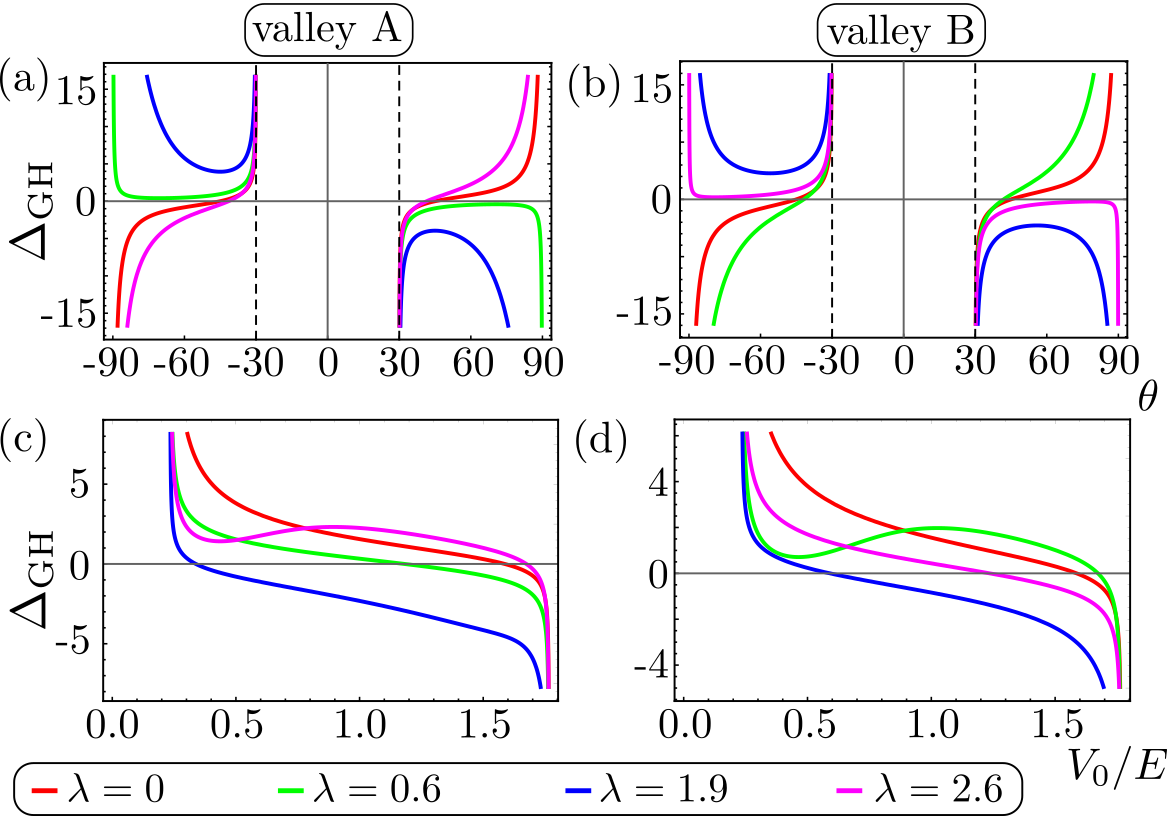}
\end{center}\vspace{-4mm}
\caption{{\bf Scalar potential:} The valley-dependent GH shift as a function of the incident angle $\theta$ for $V_0/E=1.5$ ($E=1$) for valley A in (a) and valley B in (b) at different strengths ($\lambda$) of the scalar surface potential. The critical angle $\theta_c=30^\circ$ below which no total reflection takes place. The valley-dependent GH shift as a function of $V_0/E$ for an incident angle $\theta=50^\circ$ for valley A in (c) and valley B in (d). The clean case corresponds to $\lambda=0$ (red). The other representative values of the scalar potential considered are $\lambda=0.6$ (green), $\lambda=1.9$ (blue), and $\lambda=2.6$ (magenta).}
\label{fig:fig2}
\end{figure}

For a scalar surface potential of the form $V_S=\lambda\delta(x)$, the transfer matrix that connects the two spinors across the interface is ${\cal T}=e^{-i\sigma_x\lambda}$ which results in a chirality-dependent reflection phase 
\begin{align}\label{reflecscalar}
 \phi_{r} &= -\theta -\frac{\pi}{2} + 2\tan^{-1}\zeta,~~{\rm where} \nonumber \\
 \zeta &= \frac{\tilde{\eta}\cos\theta(1-\beta_{\chi}\tan\lambda)}{2\eta\chi(\tan\lambda+\beta_{\chi})-\tilde{\eta}\sin\theta(1-\beta_{\chi}\tan\lambda)}.
\end{align}
The GH shift, in presence of a scalar potential, is chirality dependent 
\begin{align}\label{GHIF_scalar1}
\Delta_{\rm GH}&=\frac{\tilde{\kappa}^{2}(1-\tan^{2}\lambda)+2\epsilon\chi\tilde{\kappa}\tan\lambda+\epsilon\sec^{2}\lambda\cos^{2}{\bar\theta}}{E\tilde{\kappa}\sin{\bar\theta}\cos{\bar\theta}(1-\epsilon+2\tilde{\kappa}\chi\tan\lambda+(1+\epsilon)\tan^{2}\lambda)},
\end{align}
where $\sin{\theta_c}\equiv-\epsilon=V_0/E-1$ as defined before and $\tilde{\kappa}=\sqrt{\sin^2{\bar\theta}-\sin^2\theta_c}=\sqrt{\bar{k}_y^2-(E-V_0)^2}/E=\kappa/E$ for $\bar{k}_z=0$. 

\begin{figure*}
\begin{center}
  \includegraphics[width=1.0\linewidth]{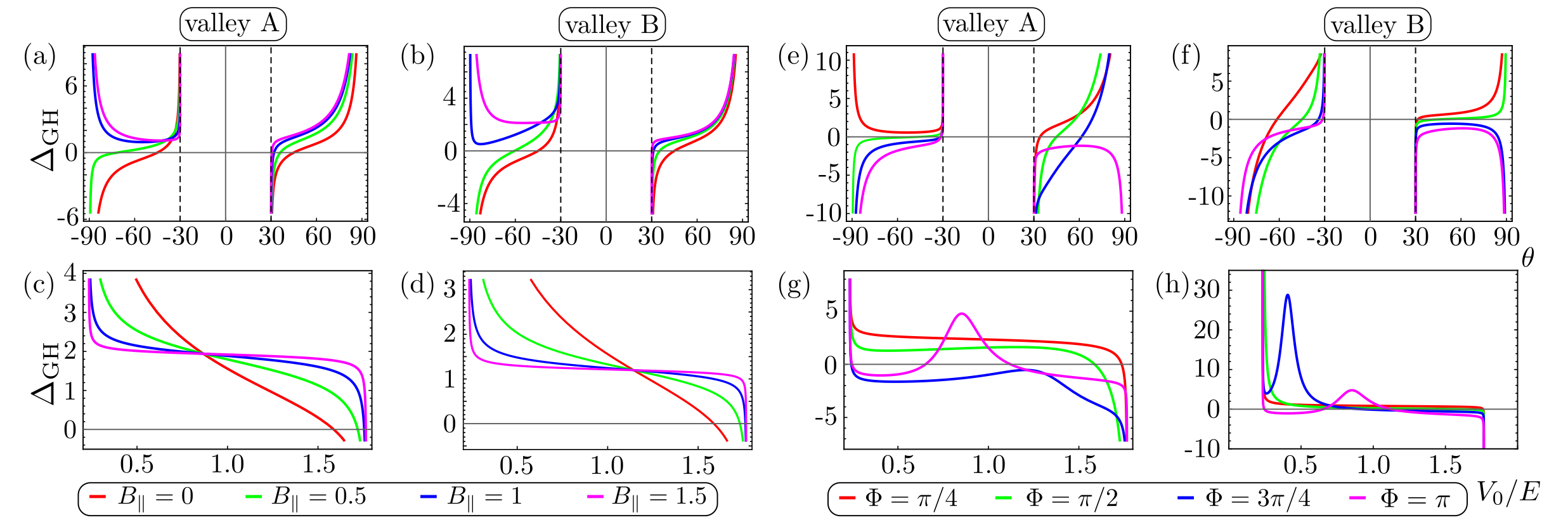}
\end{center}\vspace{-4mm}
\caption{{\bf Magnetic potential:} The valley-dependent GH shift as a function of the incident angle $\theta$ for $V_0/E=1.5$ ($E=1$) for valley A in (a) and valley B in (b) at different strengths of the in-plane magnetic field $B_\parallel$ keeping the orientation $\Phi=\pi/10$ and again for valley A in (e) and valley B in (f) at different orientations of the in-plane magnetic field $\Phi$ keeping the strength $B_\parallel=1$. The critical angle $\theta_c=30^\circ$ as before. The valley-dependent GH shift is plotted as a function of $V_0/E$ with an incident angle $\theta=50^\circ$ for valley A in (c) and valley B in (d) varying $B_\parallel$ which shows a plateau-like behavior as $B_\parallel$ is increased. The same varying $\Phi$ is shown in (g) for valley A and (h) for valley B that features a peak for certain values of the orientation which are distinct for different valleys. The clean case corresponds to $B_\parallel=0$ (red) in (a)-(d) and the other parameter values are shown in different colors.}
\label{fig:fig3a}
\end{figure*}

The IF shift, on the other hand, is insensitive to the scalar potential and retains its form as in Eq.~\ref{shiftclean} \ie
\begin{align}
 \Delta_{\rm IF} = -\frac{\chi}{E\tan{\bar\theta}},
 \label{GHIF_scalar2}
\end{align}
reflecting topological robustness against such type of potential. In the limit $\lambda\rightarrow 0$, we retrieve the clean results of Eq.~\ref{shiftclean}.

At any finite strength of the scalar potential $\lambda$, the GH shift is evidently valley-dependent and from the expression of the reflection phase $\phi_r$ in Eq.~\ref{reflecscalar}, $\lambda=n\pi$ ($n$ integers) has the same effect as $\lambda=0$ (clean case). Figure~\ref{fig:fig2} (a)-(b) show the functional dependence of $\Delta_{\rm GH}$ on the incident angle $\theta$ at various values of $\lambda$. Here, we fix $V_0/E=1.5$ which, in turn, gives a critical angle $\theta_c=30^\circ$ below which total reflection does not occur. The main features of Fig.~\ref{fig:fig2} (a)-(b) are highlighted below which are distinct for different chiralities (also referred to as valleys). \\

\noindent
{\bf Valley A ($\chi=+1$):} For $0\le\lambda<\lambda_1$ where $\lambda_1=\pi/6$, $\Delta_{\rm GH}$ for valley A behaves qualitatively similar to the clean case, however, the value of $\theta$ at which its changes the sign, denoted $\theta^{\ast(+)}$, is $\lambda$-dependent. We find, $\theta^{\ast(+)}=\sin^{-1}\sqrt{\alpha^2_++1/4}$, where
\begin{align}
    \alpha_+ = \frac{2 \tan \lambda +\sqrt{\left(6 \cos (2 \lambda )-\frac{1}{2} \cos (4
    \lambda )+\frac{7}{2}\right) \sec ^4(\lambda )}}{8-2 \sec
    ^2\lambda}.
\end{align}
In the aforementioned range, $\theta^{\ast(+)}$ gradually moves from $45^\circ$ (at $\lambda=0$) to $90^\circ$ (at $\lambda=\lambda_1$) monotonically. The numerator in Eq.~\ref{GHIF_scalar1} for valley A changes sign at $\bar\theta=90^\circ$ across $\lambda=\lambda_1$, by virtue of which, for $\lambda_1\le\lambda \le \lambda_2$ where $\lambda_2=2\pi/3$, $\Delta_{\rm GH}$ for valley A remains finite irrespective of the incident angle $\theta$, \ie $\theta^{\ast(+)}$ does not exist. The numerator for valley A changes sign again at $\bar\theta=90^\circ$ across $\lambda=\lambda_2$. As a result, for $\lambda_2 < \lambda\le\pi$, $\theta^{\ast(+)}$ decreases from $90^\circ$ (at $\lambda=\lambda_2$) to $45^\circ$ (at $\lambda=\pi$), however not monotonically (it obtains a minimum somewhere before $\lambda=\pi$).\\ 

\noindent
{\bf Valley B ($\chi=-1$):} For $0\le\lambda<\lambda_3$ where $\lambda_3=\pi/3$, $\Delta_{\rm GH}$ for valley B behaves qualitatively similar to the clean case, however, the value of $\theta$ at which its changes the sign, denoted $\theta^{\ast(-)}$, is $\lambda$-dependent. We find, $\theta^{\ast(-)}=\sin^{-1}\sqrt{\alpha^2_-+1/4}$, where
\begin{align}
    \alpha_- = \frac{-2 \tan \lambda +\sqrt{\left(6 \cos (2 \lambda )-\frac{1}{2} \cos (4
    \lambda )+\frac{7}{2}\right) \sec ^4(\lambda )}}{8-2 \sec
    ^2\lambda}.
\end{align}
In the aforementioned range, $\theta^{\ast(-)}$ increase from $45^\circ$ (at $\lambda=0$) to $90^\circ$ (at $\lambda=\lambda_3$) but nonmonotonically (it attains a minimum somewhere in-between). The numerator in Eq.~\ref{GHIF_scalar1} for valley B changes sign at $\bar\theta=90^\circ$ across $\lambda=\lambda_3$, by virtue of which, for $\lambda_3\le\lambda \le \lambda_4$ where $\lambda_4=5\pi/6$, $\Delta_{\rm GH}$ for valley B remains finite irrespective of the incident angle $\theta$, \ie $\theta^{\ast(-)}$ does not exist, just like valley A. The numerator for valley B changes sign again at $\bar\theta=90^\circ$ across $\lambda=\lambda_4$. Consequently, for $\lambda_4 < \lambda\le\pi$, $\theta^{\ast(-)}$ decreases monotonically from $90^\circ$ (at $\lambda=\lambda_4$) to $45^\circ$ (at $\lambda=\pi$).

In summary, a notable aspect, which gains prominence from the above analysis, is that the GH shift, in presence of scalar surface potential, can behave in vastly different ways for the two valleys -- it can vanish for one of the valleys for certain values of the strength of the potential while, at the same time, it remains finite for the other. Note the plots of $\Delta_{\rm GH}$ against $\theta$ for the two valleys overlap at the transition points $\lambda=\lambda_{1,2,3,4}$. Figure~\ref{fig:fig2} (c)-(d) display the behavior of $\Delta_{\rm GH}$ against the ratio of the barrier height to the incident energy $V_0/E$ which also feature a strong valley-dependence in presence of a scalar surface potential, namely, the values of $V_0/E$ at which the GH shift vanishes, depends on the chirality, unlike the clean case. As the IF is not influenced by the scalar potential, we do not show its parametric dependence which is already discussed in Ref.~\onlinecite{jiang2015topological}.

\subsection{Magnetic potential}

\begin{figure*}
\begin{center}
  \includegraphics[width=1.0\linewidth]{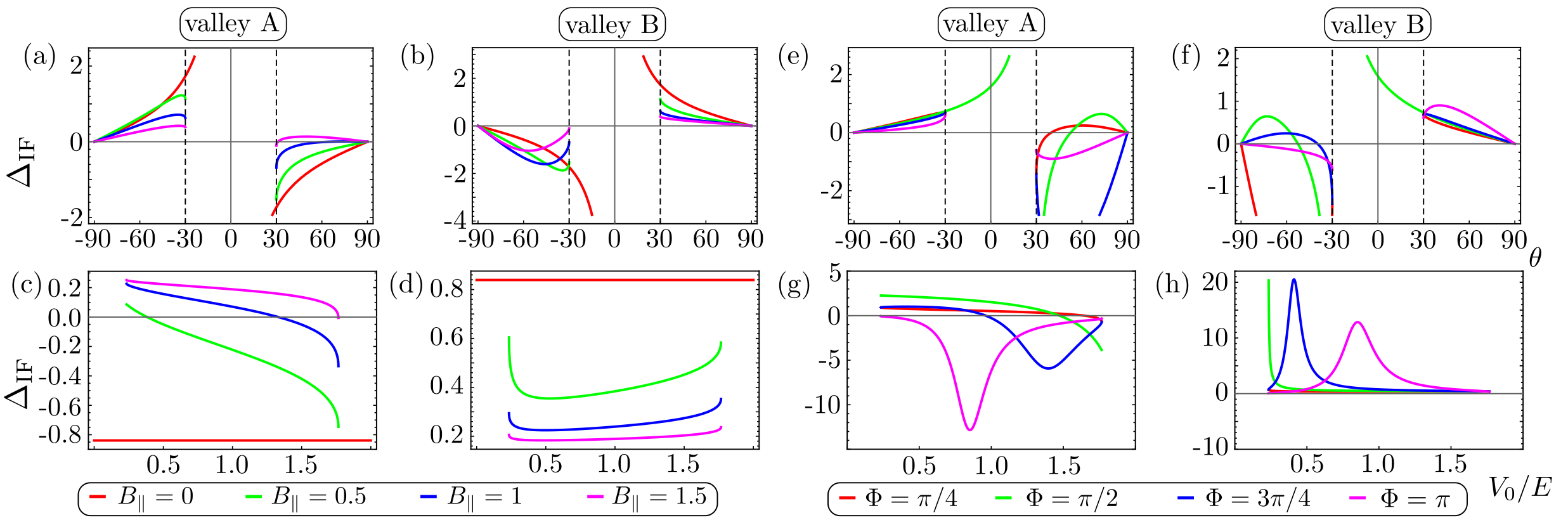}
\end{center}\vspace{-4mm}
\caption{{\bf Magnetic potential:} The valley-dependent IF shift as a function of the incident angle $\theta$ for $V_0/E=1.5$ ($E=1$) for valley A in (a) and valley B in (b) at different strengths of the in-plane magnetic field $B_\parallel$ keeping the orientation $\Phi=\pi/10$ and again for valley A in (e) and valley B in (f) at different orientations of the in-plane magnetic field $\Phi$ keeping the strength $B_\parallel=1$. The critical angle $\theta_c=30^\circ$ as before. The IF shift is plotted as a function of $V_0/E$ with an incident angle $\theta=50^\circ$ for valley A in (c) and valley B in (d) varying $B_\parallel$. A vanishing IF shift at this incidence is observed in presence of a finite $B_\parallel$ which is specific to only one of the valleys (namely, A). The same varying $\Phi$ is shown in (g) for valley A and (h) for valley B that displays a peak for certain values of the orientation which are distinct for different valleys. The clean case corresponds to $B_\parallel=0$ (red) in (a)-(d) and the other parameter values are shown in different colors.}
\label{fig:fig3b}
\end{figure*}

For a magnetic impurity present all over the interface with field orientation $\vec{ \bf B}=(B_x, B_y, B_z)$, the transfer matrix is 
\begin{align}
    {\cal T} = e^{-i\sigma_x(\vec{ \bf B}\cdot{\vec{\bf \sigma}})} = e^{-iB_x}e^{B_y\sigma_z-B_z\sigma_y}.
\end{align}
The out-of-plane component $B_x$ contributes merely as a phase factor and does not influence the shifts. For the remaining components, it is useful to introduce the parameters $B_{\parallel}=\sqrt{B_{y}^{2}+B_{z}^2}$ and $\tan\Phi=B_{z}/B_{y}$. 
In terms of these parameters, the reflection phase is 
\begin{align}\label{phasemag}
 \phi_{r} &= -\theta-\frac{\pi}{2} + 2\tan^{-1}\zeta_B,~~{\rm where} \nonumber \\
 \zeta_B &= \frac{\tilde{\eta}\cos{\theta}(b_{B}+ \beta_{\chi}a_{B})}{2\eta\chi(\beta_{\chi} +  a_{B})-\tilde{\eta}\sin\theta(b_{B}+ \beta_{\chi}a_{B})},
\end{align}
with the parameters $a_B$ and $b_B$ defined as 
\begin{align}
 a_{B} = \frac{\sin{\Phi}\tanh{B_{\parallel}}}{1+\cos{\Phi}\tanh{B_{\parallel}}}~~;~~ 
 b_{B} = \frac{1-\cos{\Phi}\tanh{B_{\parallel}}}{1+\cos{\Phi}\tanh{B_{\parallel}}}. 
\end{align}
The corresponding shifts are given by
\begin{align}
    \Delta_{\rm GH} =\frac{1}{E\tilde{\kappa}\cos{\bar\theta}}\frac{{\cal N}_{\rm GH}}{{\cal D}} ~~;~~
    \Delta_{\rm IF} =\frac{2\cos{\bar\theta}}{E}\frac{{\cal N}_{\rm IF}}{{\cal D}},
\label{GHIF_mag1}
\end{align}
where 
\begin{align}
{\cal N}_{\rm GH} &=\tilde{\kappa}^{2}(b_B^2+1-2a_B^2)+
\tilde{\kappa}(1-b_B^2)\sin{\bar\theta} \nonumber \\
&~~+2\epsilon[\chi\tilde{\kappa} a_B(1-b_B)-(a_B^2-b_B)\cos^{2}{\bar\theta}], \nonumber \\
{\cal N}_{\rm IF} &= a_B(1+b_B)\sin{\bar\theta}+a_B\tilde{\kappa}(1-b_B) \nonumber \\
&~~+\chi[(a_B^2+b_B)\epsilon+a_B^2-b_B], \nonumber \\
{\cal D} &= [(\sin{\bar\theta}+\tilde{\kappa})(1+a_B^2-2\chi a_B\sin{\bar\theta}) \nonumber \\
&+(\sin{\bar\theta}-\tilde{\kappa})(a_B^2+b_B^2-2a_Bb_B\chi\sin{\bar\theta}) \nonumber \\
&+2\epsilon(a_B\chi(1+b_B)-a_B^2\sin{\bar\theta}-b_B\sin{\bar\theta})].
\label{GHIF_mag2}
\end{align}
Eq.~\ref{GHIF_mag1} matches with the clean results in the limit $B_\parallel\rightarrow 0$. 

Figure \ref{fig:fig3a} displays the behaviour of the GH shift $\Delta_{\rm GH}$ in presence of magnetic surface potential at various values of the in-plane strength $B_\parallel$ and the orientation of the magnetic field characterized by the angle $\Phi$ defined previously. In Fig.~\ref{fig:fig3a} (a)-(d), $B_\parallel$ is being varied while keeping the orientation $\Phi$ fixed to $\pi/10$. In Fig.~\ref{fig:fig3a} (e)-(h), $B_\parallel=1$ while the orientation $\Phi$ is being varied. The out-of-plane component has no effect on the shifts and taken to be zero. In Fig.~\ref{fig:fig3a} (a)-(b), $\Delta_{\rm GH}$ is plotted against the incident angle $\theta$ for the two different valleys, given $V_0/E=1.5$. Evidently, $\Delta_{\rm GH}$ is no more an odd function of $\theta$ unlike the clean or the scalar case. The distinctive behavior of the two valleys are prominent that can be explained in the same way as the scalar potential. In Fig.~\ref{fig:fig3a} (c)-(d), $\Delta_{\rm GH}$ is plotted against $V_0/E$ which, for the both the valleys, features a plateau-like behavior as $B_\parallel$ is increased. When the orientation $\Phi$ is varied, $\Delta_{\rm GH}$ plotted against $\theta$ as in Fig.~\ref{fig:fig3a} (e)-(f) behaves in a qualitatively similar way to Fig.~\ref{fig:fig3a} (a)-(b), however, its variation against $V_0/E$ is observed to develop conspicuous peaks for certain orientations whose height can differ in order of magnitudes between the two valleys as shown in Fig.~\ref{fig:fig3a} (g)-(h).  

The behavior of the IF shift in presence of the magnetic surface potential is quite remarkable as shown in Fig.~\ref{fig:fig3b}. Similar to Fig.~\ref{fig:fig3a}, the variation with respect to $B_\parallel$ keeping $\Phi$ fixed is displayed in Fig.~\ref{fig:fig3b} (a)-(d) while the opposite is shown in Fig.~\ref{fig:fig3b} (e)-(h). Note that at any finite $B_\parallel$, the IF shift is defined only for $|\theta|>\theta_c$ where the total reflection takes place because of the parameter $\tilde{\kappa}$ unlike the clean case. For the clean case, this parameter in absent in the expression of $\Delta_{\rm IF}$ as in Eq.~\ref{shiftclean}. Besides the difference in magnitude, $\Delta_{\rm IF}$ as a function of $\theta$ can have the same sign for the two valleys depending on the values of $B_\parallel$ when it is gradually increased. In Fig.~\ref{fig:fig3b} (a)-(b), such a behavior is observed for $B_\parallel=1.5$. Further from Eq.~\ref{GHIF_mag1}, $\Delta_{\rm IF}$ can change sign for $\chi=+1$ and vanishes at a specific value of $\epsilon$ or equivalently $V_0/E$ that depends on $B_\parallel$ and $\Phi$. This is not the case for the other valley [see Fig.~\ref{fig:fig3b} (c)-(d)]. In the clean case, \ie for $B_\parallel=0$, the factor containing $\epsilon$ drops off which results in $\Delta_{\rm IF}$ being independent of $V_0/E$ with the chirality $\chi$ appearing as a prefactor. Thus, magnetic surface potential can lead to a vanishing IF shift for a specific valley ($\chi=+1$) at an incident angle $\theta_c<\theta<\pi/2$ while not having any such effect on the other. Upon varying the orientation $\Phi$ as in Fig.~\ref{fig:fig3b} (e)-(f), for both the valleys, the behavior appears similar to varying $B_\parallel$. In addition, when plotted against $V_0/E$, $\Delta_{\rm IF}$ exhibits a peak, similar to $\Delta_{\rm GH}$, for certain values of $\Phi$ as shown in Fig.~\ref{fig:fig3b} (g)-(h) for $B_\parallel=1$.  

\subsection{Rashba spin-orbit potential}

\begin{figure}[t]
\begin{center}
  \includegraphics[width=1.0\linewidth]{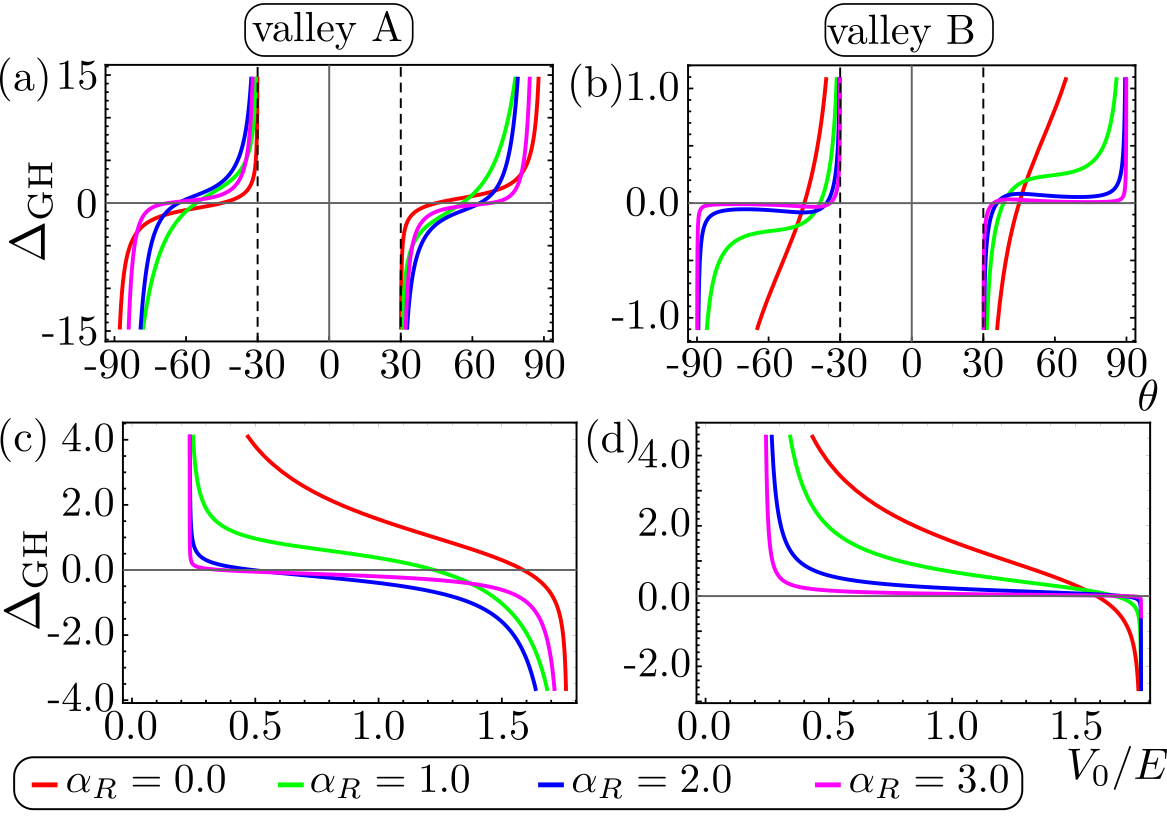}
\end{center}\vspace{-4mm}
\caption{{\bf Rashba potential:} The valley-dependent GH shift as a function of the incident angle $\theta$ for $V_0/E=1.5$ ($E=1$) for valley A in (a) and valley B in (b) at different strengths ($\alpha_R$) of the Rashba spin-orbit potential. The same as a function of $V_0/E$ for an incident angle $\theta=50^\circ$ for valley A in (c) and valley B in (d). The clean case corresponds to $\alpha_R=0$ (red).}
\label{fig:fig4}
\end{figure}

For surface potentials resulting from Rashba type spin-orbit coupling $\mathcal{V}=\alpha_{R}(k_{y}\sigma_{z}-k_{z}\sigma_{y})$, the transfer matrix is given as-
\begin{align}
 \mathcal{T}&=e^{-i\alpha_{R}\sigma_{x}(k_{y}\sigma_{z}-k_{z}\sigma_{y})} =e^{-\alpha_{R}(k_{z}\sigma_{z}+k_{y}\sigma_{y})}
\end{align}
and the reflection phase, in this case, $\phi_{r}$ turns out to be
\begin{align}\label{phaserashba}
 \phi_{r} &= -\theta-\frac{\pi}{2}+2\tan^{-1}\zeta_R,~~{\rm where} \nonumber \\
 \zeta_R &= \frac{\tilde{\eta}\cos\theta(1+\beta_{\chi}a_{R})}{2\eta\chi(\beta_{\chi} b_{R}+a_{R})-\tilde{\eta}\sin\theta(1+\beta_{\chi}a_{R})},
\end{align}
with the parameters $a_R$ and $b_R$ defined as
\begin{align}
 a_{R} = \frac{\cos\varphi\tanh{(\alpha_{R}k_{\parallel}})}{1+\sin\varphi\tanh{(\alpha_{R}k_{\parallel}})}~;~b_{R} = \frac{1-\sin\varphi\tanh{(\alpha_{R}k_{\parallel}})}{1+\sin\varphi\tanh{(\alpha_{R}k_{\parallel}})}. 
\end{align}
Here, $k_{\parallel}=\sqrt{k^2_{y}+k^2_{z}}$ and $\tan\varphi={k_z}/{k_y}$.

Note the expression of $\zeta_R$ can be obtained from that of $\zeta_B$ by identifying $B_y\leftrightarrow-\alpha_R k_z$ and $B_z\leftrightarrow\alpha_R k_y$, or equivalently, $B_\parallel\leftrightarrow\alpha_R k_\parallel$ and $\Phi\leftrightarrow\pi/2+\varphi$, which, in turn, identifies $a_B/b_B\leftrightarrow a_R$ and $b^{-1}_B\leftrightarrow b_R$. This readily yields Eq.~\ref{phaserashba} from Eq.~\ref{phasemag}. 
However, the results for the shifts that follow are significantly different from the magnetic impurity case since $a_R$ and $b_R$ are momentum-dependent parameters unlike $a_B$ and $b_B$. This has distinct effects on the two shifts when we adopt a reference by setting $\bar{k}_z=0$. 

In detail, in this setting, the Rashba type surface potential appears to have no effect on the IF shift \ie
 \begin{equation}
    \Delta_{\rm IF}=-\frac{\chi}{E\tan\bar{\theta}},
\end{equation}
same as the clean case.
The chirality-dependent GH shift, however, is modified and given by
\begin{equation}
    \Delta_{\rm GH}=\frac{1}{\kappa\cos\bar{\theta}}\frac{\mathcal{N}_{\rm R}}{\mathcal{D}_{\rm R}},
\end{equation}
where 
\begin{align}
 \mathcal{N}_{\rm R} &= (\tilde{\kappa}^{2}+\epsilon\cos^{2}\bar\theta-2\chi\alpha_{R}E\tilde{\kappa}^{2}\cos^{2}\bar\theta)\sech^{2}(\alpha_{R}E\sin\bar\theta), \nonumber \\
 \mathcal{D}_{\rm R} &= \sin{\bar\theta}(1-\epsilon)[1+\tanh^{2}({\alpha_{R}E\sin\bar\theta})] \nonumber \\
 &+2\chi\tanh({\alpha_{R}E\sin\bar\theta})(\epsilon-\sin^{2}\bar\theta).
\end{align}
It is straightforward to show that ${\cal N}_{\rm R}$ and ${\cal D}_{\rm R}$ approach their corresponding clean values in the limit $\alpha_R\rightarrow 0$.

\begin{figure}[t]
\begin{center}
  \includegraphics[width=1.0\linewidth]{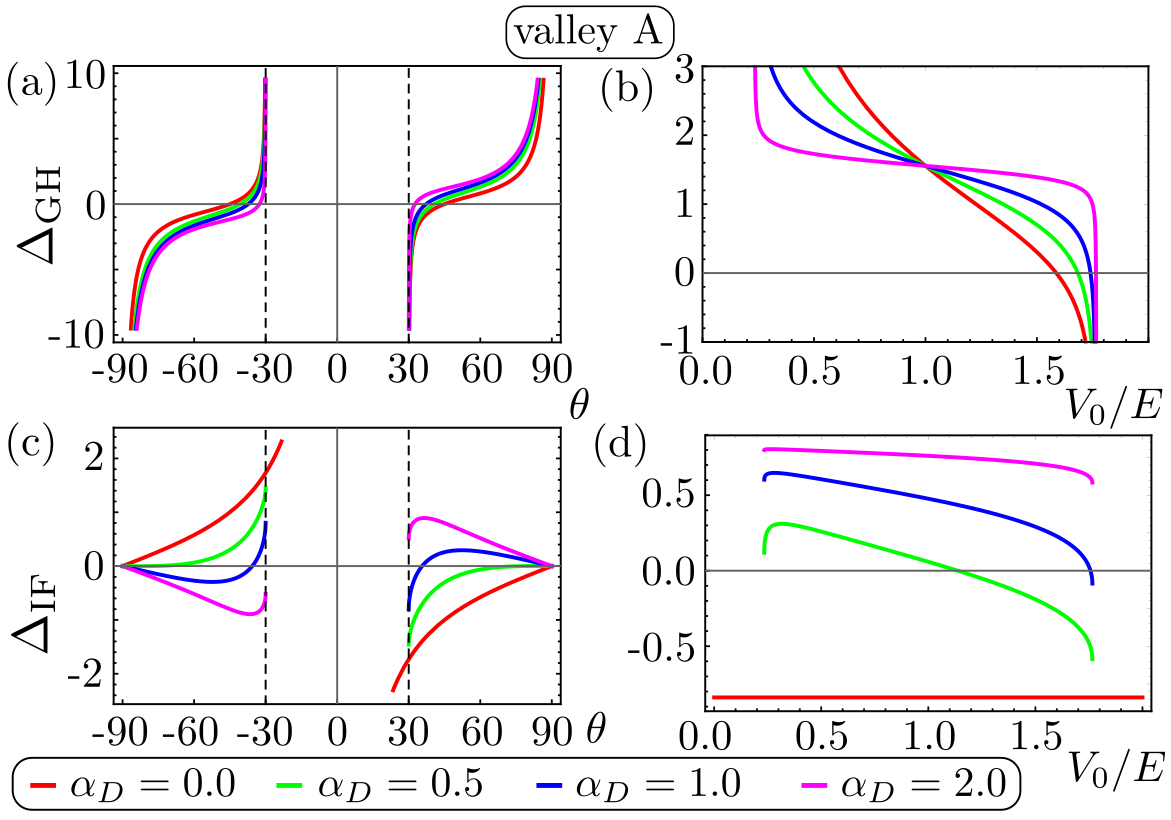}
\end{center}\vspace{-4mm}
\caption{{\bf Dresselhaus potential:} The GH shift as a function of the incident angle $\theta$ for $V_0/E=1.5$ ($E=1$) in (a) and the same as a function of $V_0/E$ with $\theta=50^\circ$ in (b) for valley A at different strengths ($\alpha_D$) of the Dresselhaus spin-orbit potential. The IF shift under the same setup in (c) and (d). Note here $\Delta_{\rm GH}$ is valley-independent while $\Delta_{\rm IF}\propto \chi$ like the clean case ($\alpha_D=0$).}
\label{fig:fig5}
\end{figure}

The influence of Rashba-type surface potential on the shifts is depicted in Fig.~\ref{fig:fig4}. The GH shift behaves qualitatively very similar to the clean case in the sense that it is odd in $\theta$ and for all values of $\alpha_R$ below $\alpha_R\sim10$ (beyond this, $\Delta_{\rm GH}\sim 10^{-5}$ and gets further suppressed with increasing $\alpha_R$), there exists a $\theta^\ast$ at which the shift vanishes (and across which it changes sign) as can be seen in Fig.~\ref{fig:fig4} (a)-(b) for both the valleys. The same comparison holds for the plots of $\Delta_{\rm GH}$ vs. $V_0/E$ [Fig.~\ref{fig:fig4} (c)-(d)] except that at large values of $\alpha_R$, the GH shift plateaus over the entire range of $V_0/E$ for both the valleys. Further, as shown above, the IF shift is not affected, and so, we do not provide any parametric plot of $\Delta_{\rm IF}$ for this case. 

\subsection{Dresselhaus spin-orbit potential}

For surface potentials that arise from Dresselhaus type spin-orbit coupling $\mathcal{V}=\alpha_{D}(k_{y}\sigma_{y}-k_{z}\sigma_{z})$, the transfer matrix is given by
\begin{align}
 \mathcal{T} = e^{-i\alpha_{D}\sigma_{x}(k_{y}\sigma_{y}-k_{z}\sigma_{z})} = e^{\alpha_{D}(k_{z}\sigma_{y}+k_{y}\sigma_{z})},
\end{align}
which results in a reflection phase
\begin{align}\label{phasedress}
 \phi_{r} &= -\theta -\frac{\pi}{2}+2\tan^{-1}\zeta_D,~~{\rm where} \nonumber \\
 \zeta_D &= \frac{\tilde{\eta}\cos\theta(b_D-\beta_{\chi}a_D)}{2\eta\chi(\beta_{\chi}-a_D)-\tilde{\eta}\sin\theta(b_D-\beta_{\chi}a_D)},
\end{align}
with the parameters $a_D$ and $b_D$ defined as
\begin{align}
a_D=\frac{\sin\varphi\tanh(\alpha_D k_{\parallel})}{1+\cos\varphi\tanh(\alpha_D k_{\parallel})}~;~b_D=\frac{1-\cos\varphi\tanh(\alpha_D k_{\parallel})}{1+\cos\varphi\tanh(\alpha_D k_{\parallel})}.
\end{align}
Here a straightforward identification $a_B\leftrightarrow-a_D$ and $b_B\leftrightarrow b_D$ (by virtue of $B_\parallel\leftrightarrow\alpha_D k_\parallel$ and $\Phi\leftrightarrow 2\pi-\varphi$) would yield Eq.~\ref{phasedress} from Eq.~\ref{phasemag}.

Following analogous calculations as the Rashba case above, we obtain the shifts as
\begin{align}
\Delta_{\rm GH} = \frac{1}{\kappa\cos\bar{\theta}} \frac{\mathcal{N}_{\rm D}^{(\rm GH)}}{{\cal D}_{\rm D}}~~;~~ \Delta_{\rm IF} = -\frac{\chi}{E\tan\bar{\theta}}\frac{\mathcal{N}_{\rm D}^{(\rm IF)}}{{\cal D}_{\rm D}},
\end{align}
where
\begin{align}
 \mathcal{N}_{\rm D}^{(\rm IF)} &= (1-\epsilon)[1-\tanh^2(\alpha_D E\sin\bar{\theta})]\sin\bar{\theta}\nonumber \\
 &- 2\tilde{\kappa}\tanh(\alpha_D E\sin\bar{\theta})-2\tanh^2(\alpha_D E\sin\bar{\theta})\sin\bar{\theta}, \nonumber \\
 \mathcal{N}_{\rm D}^{(\rm GH)} &= (\tilde{\kappa}^{2}+\epsilon\cos^{2}\bar{\theta}+2\alpha_D\epsilon\tilde{\kappa}E\cos^{2}\bar{\theta}) \nonumber \\ 
 &+ 2\tanh(\alpha_D E\sin\bar{\theta})\tilde{\kappa}\sin\bar{\theta}  \nonumber \\ 
 &+ \tanh^{2}(\alpha_D E\sin\bar{\theta})[\tilde{\kappa}^{2}-\epsilon\cos^{2}\bar{\theta}-2\alpha_D\epsilon\tilde{\kappa}E\cos^{2}\bar{\theta}],
\end{align}
and
\begin{align}
 \mathcal{D}_{\rm D} &= \sin\bar{\theta}(1-\epsilon)+2\tilde{\kappa}\tanh(\alpha_D E\sin\bar{\theta}) \nonumber \\
 &+ \tanh^2(\alpha_D E\sin\bar{\theta})\sin\bar{\theta}(1+\epsilon).
\end{align}
Again, the clean results are readily obtained in the limit $\alpha_D\rightarrow 0$.

The Dresselhaus-type surface potential has quite a few exotic effects on both the shifts. Firstly, from the above expressions, $\Delta_{\rm GH}$ is chirality-independent and odd in $\theta$ while $\Delta_{\rm IF}$ carries the same chirality dependence as the clean case \ie $\Delta_{\rm IF}\propto \chi$. This is reflected in Fig.~\ref{fig:fig5}. In Fig.~\ref{fig:fig5} (a), $\Delta_{\rm GH}$ is plotted against $\theta$ keeping $V_0/E=1.5$. Similar to the other potentials, a $\theta^\ast$ exists at which $\Delta_{\rm GH}$ vanishes in a given range of $\alpha_D$ beyond which $|\Delta_{\rm GH}|$ remains finite. In Fig.~\ref{fig:fig5} (b), $\Delta_{\rm GH}$ is plotted against $V_0/E$ keeping $\theta=50^\circ$. We note while at small values of $\alpha_D$, the behavior of $\Delta_{\rm GH}$ looks qualitatively similar to the clean case, as $\alpha_D$ is gradually increased, the plateauing effect starts dominating and eventually at very large values of $\alpha_D$, $\Delta_{\rm GH}$ saturates at a value that increases with $\theta$.

For the IF shift, we find that the Dresselhaus type surface potential can interestingly lead to phenomenon like valley inversion as seen in Fig.~\ref{fig:fig5} (c) in the following manner. Let us denote the value of $\theta$ at which the IF shift vanishes by $\theta_{\rm IF}^\ast$. For the clean case, $|\theta_{\rm IF}^\ast|=\pi/2$. When $\alpha_D$ is increased, there appears another $\theta^\ast_{\rm IF}$ such that $|\theta^\ast_{\rm IF}|<\pi/2$ which gradually approaches $\theta_c$ as $\alpha_D$ approaches
\begin{align}
    \alpha_D^\ast = \frac{1}{E\sin\theta_c}\tanh^{-1}\frac{V_0}{V_0+2}.
\end{align}
If $\alpha_D$ is increased further beyond this value, a valley inversion takes place as $\Delta_{\rm IF}$ changes sign across $\alpha_D=\alpha_D^\ast$. Similar phenomenon is also observed in the plot of $\Delta_{\rm IF}$ against $V_0/E$ as in Fig.~\ref{fig:fig5} (d) where the valley inversion manifests as reordering of the curves of $\Delta_{\rm IF}$ against $V_0/E$ at different values of $\theta$ as $\alpha_D$ is increased and consequently, $\Delta_{\rm IF}(\theta)\rightarrow-\Delta_{\rm IF}(\theta)$.  

\begin{figure}[t]
\begin{center}
  \includegraphics[width=1.0\linewidth]{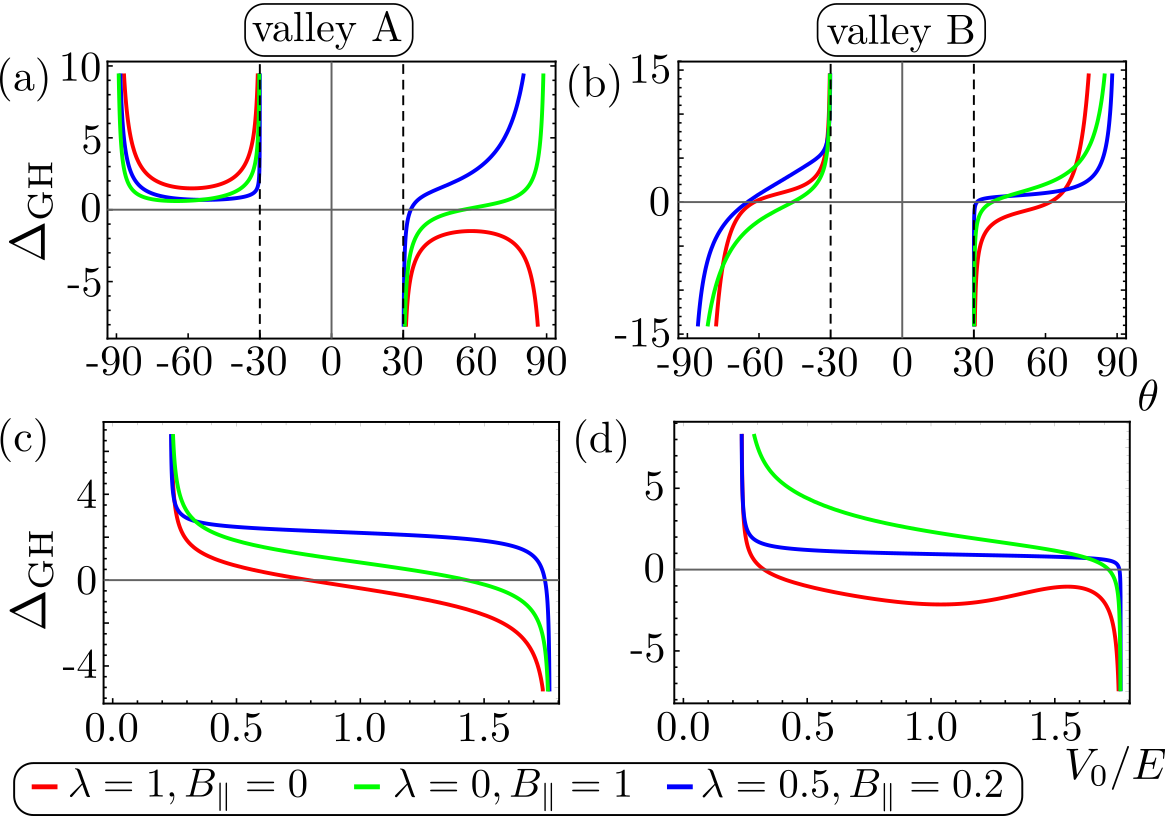}
\end{center}\vspace{-4mm}
\caption{{\bf Combination of scalar and Zeeman potential:} GH shift as a function of the incident angle $\theta$ for $V_0/E=1.5$ for valley A in (a) and valley B in (b) at different strengths of the scalar surface potential ($\lambda$) and magnetic potential ($B_\parallel$). The same as a function of $V_0/E$ at $\theta=50^\circ$ for valley A in (c) and valley B in (d). The combinations of $\lambda$ and $B_\parallel$ are specified in different colors as $\lambda=1, B_\parallel=0$ in red, $\lambda=0, B_\parallel=1$ in green, $\lambda=0.5, B_\parallel=0.2$ in blue.}
\label{fig:fig6}
\end{figure}

\section{Discussion}\label{secfive}

In this article, we revisit the phenomena of GH and IF shifts, the lateral shifts of an incident beam upon total reflection, in a Weyl semimetal system to discuss the effects of surface potentials on these shifts which could be probed in real materials. Earlier these shift were studied for a clean Weyl surface and argued that the IF shift is topological in nature, namely, it is chirality-dependent and so, can be exploited in experiments to characterize Weyl systems. The GH shift in Weyl semimetals, on the other hand, does not have such a feature. However, as we reveal in this article, this is not entirely true when the concerned surface harbors various kinds of surface potentials as is the case in real materials exhibiting Weyl nodes. Among the key observations, presence of any type of surface potential renders the GH shift strongly chirality-dependent. In fact, strong scalar or magnetic potential can yield a situation in which the GH shift remains finite irrespective of the incident angle for the allowed values of the magnitude of the chemical potential barrier ($V_0$) that distinguishes the surface. For other types of surface potentials such as of the Rashba or Dresselhaus type, the GH shift shows conspicuous departure from the clean results. When plotted as a function of the barrier height $V_0$, strong surface potentials give rise to plateauing effect in the GH shift and sometimes, peaks at specific values of $V_0$ for magnetic impurities. These effects can be particularly useful in probing magnetic Weyl systems that have recently nucleated experimental activities. 

For the IF shift, the scalar surface potentials turn out to have no effect. However, the other kinds of surface potentials do leave remarkable signatures compared to the clean case similar to when intervalley scattering takes place. Most promising are the magnetic and Dresselhaus-type impurities, in presence of which, the IF shift, which otherwise, is independent of $V_0$ and changes sign between the two valleys for the clean case, develops a strong valley asymmetry beyond a simple sign inversion and also a parametric dependence on $V_0$ such that it can vanish at certain values of $V_0$ depending on the incident angle $\theta$. Similarly, in distinction to the clean case where the IF shift vanishes only at $\theta=\pm\pi/2$, the impurities can enforce intermediate values of $0<|\theta|<\pi/2$ at which the IF shift is nullified and that too can be valley-dependent. This can potentially mask the IF shift to be identified as a topological effect in realistic Weyl systems as their surfaces would typically host various impurities even including a mixed nature. 

To investigate the effects of such mixed impurities, we study the case where the surface potential includes both the scalar and magnetic contributions as ${\cal V}=\lambda+{\bf B}\cdot\sigma$. The results are summarized in Fig.~\ref{fig:fig6}. While the IF shift (not shown) is dictated by the magnetic contribution, the GH shift is affected by both. In fact, as revealed by Fig.~\ref{fig:fig6}, the plateauing effect is stronger when both types of surface potentials are present with moderate strengths, compared to their individual influences at higher strengths. If the two impurities form distinct domains on the surface, we expect the resultant shifts to be an weighted average of the individual shifts as $\langle\Delta\rangle=w\langle\Delta_1\rangle+(1-w)\langle\Delta_2\rangle$ where the two kinds of impurities are denoted by 1 and 2 and $\langle\dots\rangle$ implies ensemble averaging that includes the impurities with probabilities $w$ and $(1-w)$ respectively. This extends to multiple types of impurities as well. 

In summary, our work extends the phenomena of GH and IF shift in Weyl semimetals beyond a clean surface and accommodates surface potentials to unveil novel features of the shifts. Our observations would provide useful guidance to experiments that are tuned to characterize Weyl systems based on such phenomena that have already found potential relevance in device applications engaging other electronic systems. Studying similar effects on other transport properties of Weyl semimetals will be addressed elsewhere. 

\begin{acknowledgements} 
NKD and KR gratefully acknowledge useful discussion with Qing-Dong Jiang and Sourin Das and thank them for critically reviewing the manuscript. KR also thanks the sponsorship, in part, by the Swedish Research Council.
\end{acknowledgements}

\bibliographystyle{apsrev}
\bibliography{references}
\end{document}